# Anomalous spectral weight transfer in the nematic state of iron-selenide superconductor[*]

C. Cai(蔡淙)[1], T. T. Han(韩婷婷)[1], Z. G. Wang(王政国)[1], L. Chen(陈磊)[1], Y. D. Wang(王宇迪)[1], Z. M. Xin(信子鸣)[1], M. W. Ma(马明伟)[1], Yuan Li(李源)[1,2], and Y. Zhang(张焱)[1,2,†]

[1]*International Center for Quantum Materials, School of Physics, Peking University, Beijing 100871, China*
[2]*Collaborative Innovation Center of Quantum Matter, Beijing 100871, China*

Nematic phase intertwines closely with high-$T_c$ superconductivity in iron-based superconductors. Its mechanism, which is closely related to the pairing mechanism of superconductivity, still remains controversial. Comprehensive characterization of how the electronic state reconstructs in the nematic phase is thus crucial. However, most experiments focus only on the reconstruction of band dispersions. Another important characteristic of electronic state, the spectral weight, has not been studied in details so far. Here, we studied the spectral weight transfer in the nematic phase of FeSe$_{0.9}$S$_{0.1}$ using angle-resolved photoemission spectroscopy and *in-situ* detwinning technique. There are two elliptical electron pockets overlapping with each other orthogonally at the Brillouin zone corner. We found that, upon cooling, one electron pocket loses spectral weight and fades away, while the other electron pocket gains spectral weight and becomes pronounced. Our results show that the symmetry breaking of electronic state is manifested by not only the anisotropic band dispersion but also the band-selective modulation of spectral weight. Our observation completes our understanding of the nematic electronic state, and put strong constraints on the theoretical models. It further provide crucial clues to understand the gap anisotropy and orbital-selective pairing in iron-selenide superconductors.

**Keywords:** nematic order, superconductivity, iron-based superconductors, angle-resolved photoemission spectroscopy

**PACS:** 74.25.Jb,74.70.Xa,79.60.-i,73.21.Ac

## 1. Introduction

Nematic phase, where a system translationally invariant but breaks rotational symmetry, attracts great interests recently due to its close intertwining with superconductivity in iron-based superconductors (FeSCs) [1-4]. The superconducting transition temperature ($T_c$) reaches optimal when the nematic transition temperature ($T_{nem}$) is suppressed by either chemical doping or pressure. It is therefore important to understand the underlying mechanism of the nematic phase, whose origin is closely related to the pairing interaction of high-$T_c$ superconductivity. Theoretically, on one hand, the nematic phase could be driven by a $d_{xz}/d_{yz}$ ferro-orbital ordering. The occupation difference between the $d_{xz}$ and $d_{yz}$ orbitals breaks the C4 rotational symmetry [5-7]. More complex orbital orderings have also been proposed [8, 9]. On the other hand, the spin fluctuation could breaks the C4 rotational symmetry spontaneously, resulting in a spin-nematic phase. It drives the nematic phase transition through a coupling among spin, charge, and lattice degrees of freedom [10, 11].

[*] Project supported by the National Natural Science Foundation of China (Grant No. 11888101, No. 91421107, and No. 11574004), the National Key Research and Development Program of China (Grant No. 2016YFA0301003 and No. 2018YFA0305602)
[†] Corresponding author. E-mail: yzhang85@pku.edu.cn

To understand the mechanism of the nematic phase, it is important to characterize experimentally how the electronic state reconstructs in the nematic phase. Angle-resolved photoemission spectroscopy (ARPES) is a powerful technique that can probe the electronic structure of materials in momentum space. Band dispersions and spectral weight distributions of the electronic state can be obtained. For the nematic phase, the reconstruction of band dispersions has been intensively studied by ARPES [12-16]. Orbital-dependent energy shifts of bands and anisotropic band dispersions have been observed. However, another important characteristic of the electronic state, the spectral weight, has not been studied in details so far. It is not clear how the spectral weights of bands evolve with temperature and whether the spectral weight distribution breaks the rotational symmetry or not in the nematic phase.

FeSe is an ideal system to study the nematic phase [17]. The $T_{nem}$ of FeSe was found to be 90 K and there is no long-range magnetic order at low temperature. Moreover, the $T_{nem}$ can be further suppressed by the sulfur substitution in $FeSe_{1-x}S_x$ [18]. With lower transition temperature, the thermal broadening effect can be suppressed and the detailed evolution of electronic state across the phase transition can be obtained. Here, we studied the electronic structure reconstruction in the nematic phase of $FeSe_{0.9}S_{0.1}$ using ARPES. We found that, the nematic phase transition not only results in an anisotropic band dispersion, but also leads to a band-selective spectral weight transfer on the electron pockets at the Brillouin zone corner (M). Upon cooling, one electron pocket loses spectral weight and eventually fades away, while the other electron pocket gains spectral weight. Such spectral weight transfer breaks the C4 rotational symmetry of Fermi surface, leaving only one anisotropic electron pocket at the M point at lowest temperature. Our results complete our understanding of the electronic structure reconstruction in the nematic phase, and put strong constrains on the theoretical models. It further provides insights into the orbital-selective pairing of iron-selenide superconductors.

2. **Materials and methods**
High quality $FeSe_{0.9}S_{0.1}$ samples were grown using the chemical vapor transport (CVT) method [18]. The $T_{nem}$ was confirmed to be ~75 K according to the resistivity and magnetic susceptibility measurements. ARPES measurements were performed using a DA30L analyzer and a helium discharging lamp. The photon energy is 21.2 eV. The overall energy resolution was ~10meV and the angular resolution was ~0.3 °. All the samples were cleaved and measured in ultrahigh vacuum with a base pressure better than $6\times10^{-11}$ mbar. To obtain photoemission signals from one single domain, piezoelectric (PZT) stacks are used to detwin the sample *in-situ* [19, 20]. No sample deformation was observed during the sample detwinning process. Note that, the PZT extension increases linearly with the PZT temperature. In order to keep a constant strain or tension on the sample, we carefully adjusted the PZT voltage during the temperature-dependent experiment.

## 3. Results and discussion

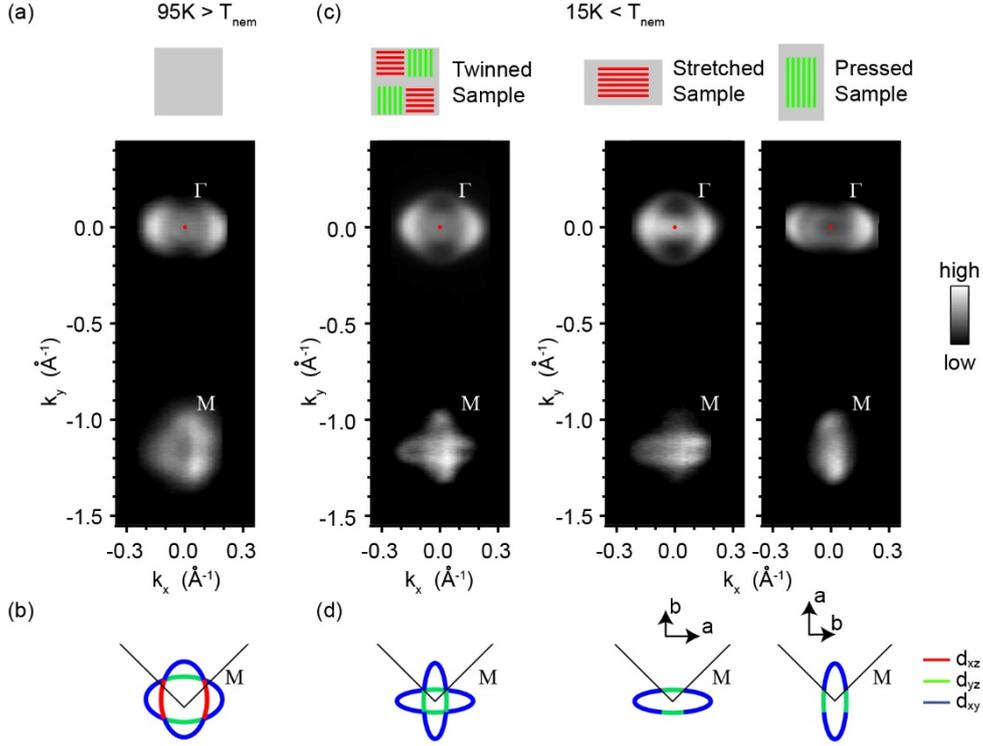

**Fig. 1.** (a) Fermi surface mapping taken at 95 K above the $T_{nem}$. (b) Schematic drawing of the Fermi surface around the M point for the data shown in panel a. (c) Fermi surface mappings taken at 15 K in twinned sample (left panel), stretched sample (middle panel), and pressed sample (right panel). (d) Schematic drawing of the Fermi surface around the M point for the data shown in panel c.

The Fermi surface of iron-based superconductors consists of hole pockets at the Brillouin zone center (Γ) and electron pockets at the Brillouin zone corner (M), respectively [21]. As shown in Figs. 1(a) and 1(b), the Fermi surface of FeSe$_{0.9}$S$_{0.1}$ consists of two elliptical electron pockets at the M point. They overlap orthogonally with each other, and are constructed by the $d_{xz}$, $d_{yz}$, and $d_{xy}$ orbitals [12, 13, 21]. In the nematic phase, the sample is twinned with domains along two perpendicular directions. By applying a uniaxial pressure or strain, the sample can be detwinned and photoemission signals from one single domain can be obtained [22, 23]. As shown in Figs 1(c) and 1(d), it is obvious that the horizontal and vertical elliptical electron pockets originate from two different domains. They show up separately in the stretched and pressed samples. The contrast between the Fermi surface mappings taken in the stretched and pressed samples indicates that the sample is fully detwinned in our experiments. In a perfectly detwinned sample, the Fermi surface consists of only one electron pocket at the M point, whose longer axis aligns with the longer axis of lattice. Here, we define the longer axis of lattice as the *x* axis. The orbital characters of the remaining electron pocket could be then attributed to the $d_{yz}$ and $d_{xy}$ orbitals [12, 13].

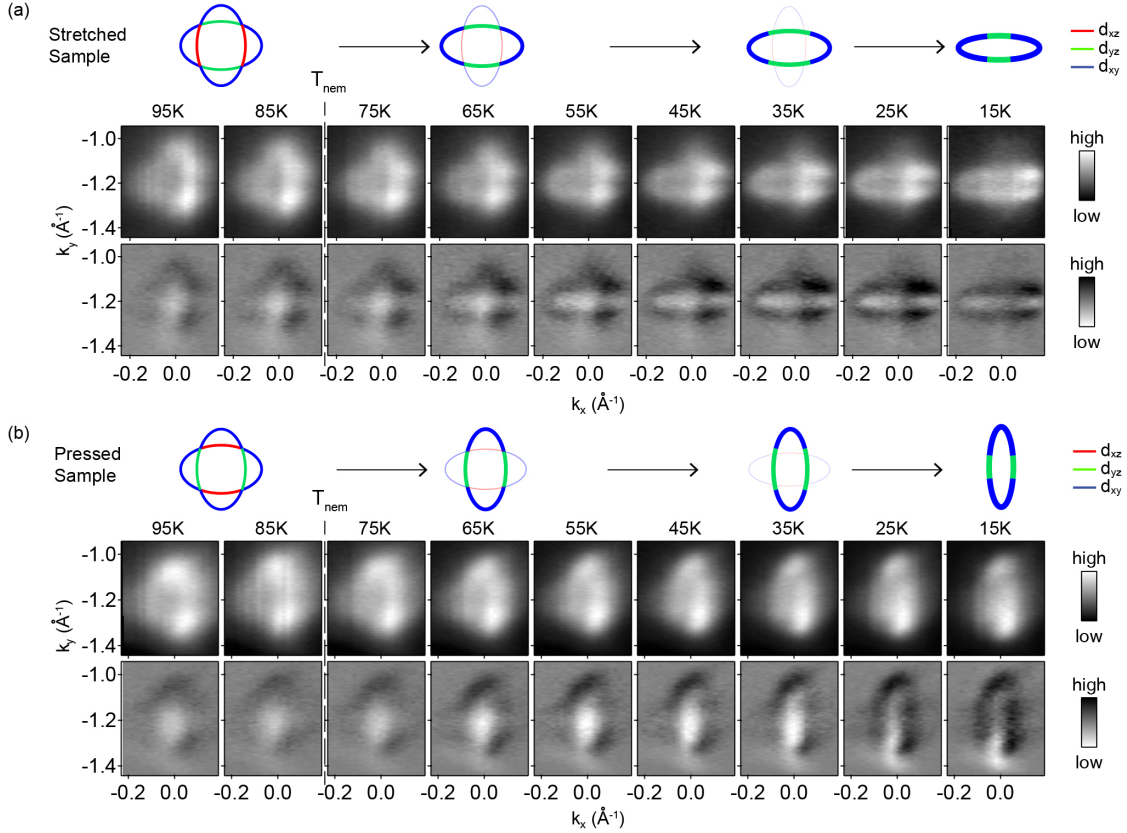

**Fig. 2**. (a) Temperature dependence of the Fermi surface mapping taken in the stretched sample. The upper panels show the raw data and the lower panels show the second derivative images. The derivation is calculated in the energy direction. (b) is the same as panel a but taken in the pressed sample. The PZT voltage was carefully adjusted to keep a constant strain or tension during the temperature dependent experiments.

To understand how the Fermi surface evolves with temperature, we took the detailed temperature dependences of Fermi surface mapping in both the stretched and pressed samples. The data are shown in Fig.2. In the stretched sample, the Fermi surface consists of two elliptical electron pockets at 95 K. When the temperature decreases below 75 K, the vertical elliptical electron pocket start to lose spectral weight and eventually fades away at 15 K. In contrast, the horizontal elliptical electron pocket start to gain spectral weight at 75 K and becomes pronounced at the lowest temperature. Similar temperature dependence was also observed in the pressed sample with all observations rotating 90 ° [Fig. 2(b)]. The consistency between the data taken in the stretched and pressed samples confirms that the observed spectral weight transfer is an intrinsic property of the nematic phase. Moreover, the size and shape of the electron pockets change little with temperature, which indicates that the spectral weight redistribution cannot be explained by any energy shift of band that has been observed in previous ARPES experiments. It should be noted that, our observation cannot be explained by a domain mixing at high temperature. First, upon warming, the nematic order parameter decreases and the nematic susceptibility increases. This suggests that the sample detwining is easier at high temperature than at low temperature with the same sample strain or tension level. Second, as shown in the following figures, the anisotropic band dispersions confirm that the sample remains detwinned during the temperature-dependent experiment.

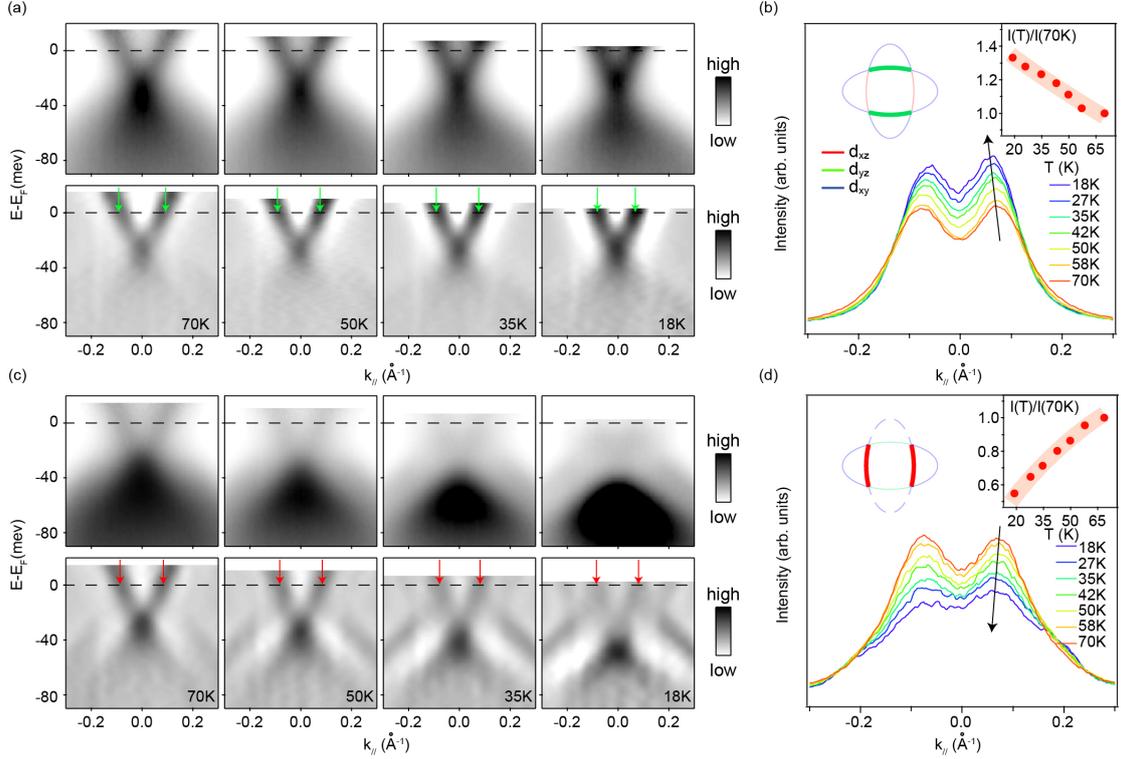

**Fig. 3**. (a) Temperature dependence of the $d_{yz}$ electron band. The upper and lower panels show the raw data and second derivative images, respectively. The derivation is calculated in the momentum direction. The Fermi crossings (k$_{FS}$) of the $d_{yz}$ electron band are illustrated using the green arrows. (b) Momentum distribution curves (MDCs) taken at the Fermi energy ($E_F$) for the data shown in panel a. Inset panel shows the temperature dependence of the spectral weight of $d_{yz}$ electron band calculated using a standard peak fitting process. (c) and (d) are the same as panel a and b, but taken on the $d_{xz}$ electron band.

The electron pockets are constructed by the $d_{xz}$, $d_{yz}$ and $d_{xy}$ electron bands. By changing both the domain direction and the experimental orientation, we could probe different electron bands selectively [23, 24]. Fig. 3 shows the temperature dependence of the $d_{yz}$ and $d_{xz}$ electron bands. The contrast between them is obvious. Upon cooling, the spectral weight of the $d_{yz}$ electron band increases while the spectral weight of the $d_{xz}$ electron band decreases. Such gain or loss of spectral weight occurs in a wide energy range that covers most of the $d_{yz}$ and $d_{xz}$ electron bands. Meanwhile, the Fermi crossings (k$_{FS}$) of the $d_{yz}$ and $d_{xz}$ electron bands shrink slightly upon cooling. The second derivative images and MDCs clearly show that both the $d_{xz}$ and $d_{yz}$ electron bands remains below $E_F$ even at the lowest experimental temperature. Note that, the strong contrast between the band dispersions taken along a and b directions is obvious for all the measured temperatures under T$_{nem}$ [Figs. 3(a) and 3(c)], which suggests that there is no domain mixing and the sample remains detwinned during the temperature-dependent experiment.

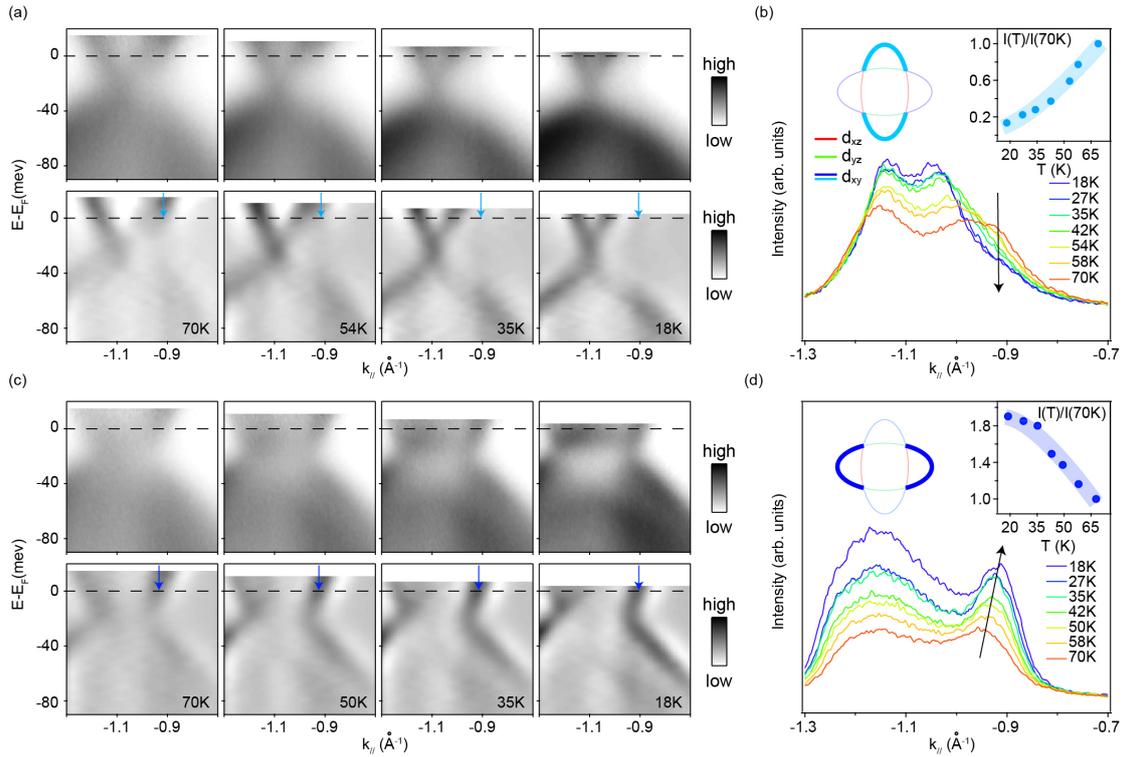

**Fig. 4.** (a) Temperature dependence of the $d_{xy}$ electron band taken along the vertical direction. The upper and lower panels show the raw data and second derivative images, respectively. The derivation is calculated in the momentum direction. The k$_{FS}$ of the $d_{xy}$ electron band are illustrated using the blue arrows. (b) Momentum distribution curves (MDCs) taken at $E_F$ for the data shown in panel a. Inset panel shows the temperature dependence of the spectral weight of $d_{xy}$ electron band calculated using a standard peak fitting process. (c) and (d) are the same as panels a and b, but taken on the $d_{xy}$ electron band along the horizontal direction.

Similar temperature dependent behaviors were also observed on the $d_{xy}$ electron bands. The $d_{xy}$ electron bands disperse along both the vertical and horizontal directions. As shown in Fig.4, upon cooling, the spectral weight of the $d_{xy}$ electron band decreases in the vertical direction, but increase in the horizontal direction. The k$_{FS}$ of the $d_{xy}$ electron band expands slightly in the horizontal direction upon cooling. All the temperature dependences of the electron bands are consistent with the temperature evolution of

Fermi surface (Fig. 2). While the energy shifts of the $d_{yz}$ and $d_{xy}$ bands result in an enhancement of electron pocket ellipticity, the spectral weight loss or gain occurs in a wide energy range and cannot be explained by any energy shift of band. Therefore, we conclude that the C4 rotational symmetry breaking not only results in the anisotropic band dispersion that has been observed in previous ARPES studies, but also leads to an undiscovered spectral weight transfer on the electron pockets.

Such spectral weight transfer has not been observed in previous ARPES experiments. One important reason is that the $T_{nem}$s are high in most iron-based superconductors. As a result, the energy reconstruction of bands is so pronounced, which conceals the spectral weight modulation. For example, in a recent ARPES study on the detwinned FeSe [13], the entire $d_{xz}$ electron band was found to be pushed above $E_F$ in the nematic phase. Therefore, it is difficult to observe the spectral weight loss of the $d_{xz}$ electron band. Moreover, in most iron-based superconductors, the reconstruction of Fermi surface is so strong that tracking the spectral weights of all the bands is very difficult. Here, in FeSe$_{0.9}$S$_{0.1}$, the energy scale of band shift is reduced and the shape and size of the electron pockets change little with temperature. Therefore, quantitate analyses of spectral weight on all electron bands could be achieved.

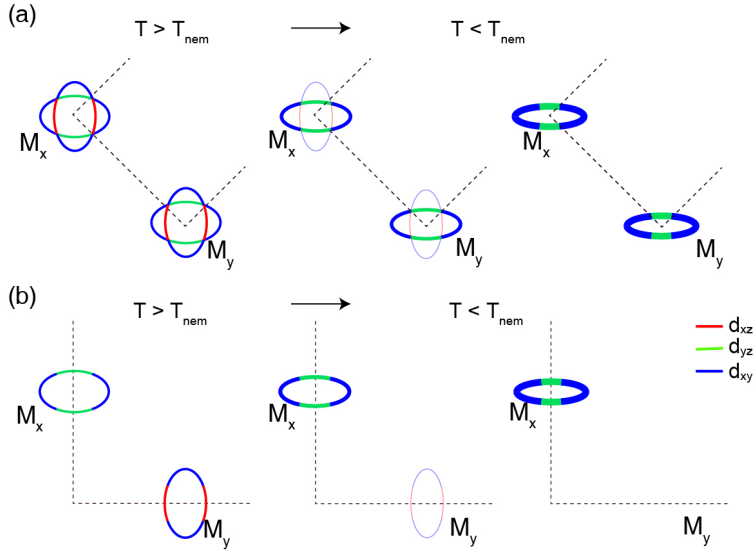

**Fig. 5.** (a) Schematic drawing of the temperature evolution of electron pockets in two-Fe Brillouin zone. (b) Schematic drawing of the temperature evolution of electron pockets in one-Fe Brillouin zone. The hole pockets at the Γ point are not shown here.

Figure 5 summarize the spectral weight transfer of the electron pockets observed here. Note that, the spectral weight transfer breaks the C4 rotational symmetry and thereby is closely related to the nematic phase transition. Moreover, the spectral weight transfer is band-selective not orbital-selective. Therefore, the spectral weight transfer cannot be explained by an orbital-selective Mott transition [25, 26]. To discuss the origin of the observed spectra weight transfer, we consider two different transfer scenarios. On one hand, the spectral weight could transfer between two elliptical electron pockets at different momenta. For instance, a complex orbital ordering could redistribute the $d_{xz}$, $d_{yz}$, and $d_{yz}$ orbital weights on the electron pockets. The $d_{xy}$ orbital weight transfers from one electron pocket to the other upon cooling, leading to the existence of one nearly pure $d_{xy}$ electron pocket and another $d_{xz}/d_{yz}$ electron pocket at the lowest temperature. Because the photoemission intensity of the $d_{xy}$ bands is normal much weaker than that of the $d_{xz}/d_{yz}$ bands [27]. Such orbital weight transfer between two electron pockets can explain the spectral weight transfer observed here. On the other hand, the spectral weight could transfer in the energy direction. In correlated materials, the coherent spectral weight of quasi-particle would transfer into an incoherent spectral weight when many-body interactions increase, and vice versa. Under this scenario, the loss and gain of spectral weight of two electron pockets may originate from their different coherencies. For instance, in the nematic phase, one electron pocket becomes coherent and thus gains spectral weight near $E_F$, while the other electron pocket becomes incoherent and thus loses spectral weight. Such band-selective change of many-body interactions may originated from the spin nematic phase. The spin fluctuation becomes anisotropic in the spin nematic phase [28], which may modulate the coherencies of electrons at $(0, \pi)$ and $(\pi, 0)$ unequally.

Our results characterize the detailed temperature evolution of Fermi surface in the nematic phase. We show that not only the shape of electron pockets but also their spectral weights become anisotropic. Such reconstruction of Fermi surface could affect the pairing interaction of superconductivity. It is believed that the superconducting pairing is mediated by the inter-pocket scattering between the hole and electron pockets. The unbalanced orbital weight distributions or coherencies of the electron pockets would result in a pairing interaction that is strongly anisotropic and orbital-selective. This is consistent with previous ARPES and scanning tunnel microscopy studies showing that the superconducting gap is strongly anisotropy and orbital-dependent in FeSe [29, 30]. Our results support the existence of orbital-selective pairing in FeSe, and provides crucial clues for understanding the pairing mechanism of the nematic superconducting phase.

4. **Conclusion and perspectives**
In summary, we studied the reconstruction of electronic structure in the nematic phase of FeSe$_{0.9}$S$_{0.1}$. The detailed temperature evolutions of spectral weight were measured for all electron bands at the M point. We found that the rotational symmetry breaking is manifested by not only an anisotropic band dispersion but also a band-selective modulation of spectral weight. One electron pocket loses spectral weight and the Fermi surface consists of only one anisotropic electron pocket at the M point in the nematic phase. To fully understand the observed spectral weight transfer on the electron pockets, further experimental and theoretical studies are required. Our results complete our understanding of the symmetry-broken electronic state in the nematic phase, and also put strong constrains on constructing the theoretical modes for both the nematic and superconducting phases in iron-based superconductors.

**Acknowledgment**

We thank Dr. Z. Liu for helpful discussion.